\documentclass[preprint,tightenlines,superscriptaddress,prb,nofootinbib,eqsecnum]{revtex4-1}
\usepackage[pdftex]{graphicx}
\usepackage{subfigure}
\usepackage{amstext}
\usepackage{amsmath}
\usepackage{endnotes}
\let\footnote=\endnote
\usepackage{amssymb}
\def\ee{{\rm e}}

\def\_#1{^{}_{#1}}

\def\be{\begin{equation}}
\def\ee{\end{equation}}
\def\bea{\begin{eqnarray}}
\def\eea{\end{eqnarray}}
\usepackage{enumerate}

\begin{document}

\title{Elliptical-like orbits on a warped spandex fabric: A theoretical/experimental undergraduate research project}

\author{Chad A. Middleton}
\email{chmiddle@coloradomesa.edu}
\affiliation{Department of Physical and Environmental Sciences, Colorado Mesa University, Grand Junction, CO 81501, U.S.A.}

\author{Dannyl Weller}
\email{drweller@mavs.coloradomesa.edu}
\affiliation{Department of Physical and Environmental Sciences, Colorado Mesa University, Grand Junction, CO 81501, U.S.A.}

\begin{abstract}
We present a theoretical and experimental analysis of the elliptical-like orbits of a marble rolling on a warped spandex fabric.   We arrive at an expression describing the angular separation between successive apocenters, or equivalently successive pericenters, in both the small and large slope regimes.  We find that a minimal angular separation of $\sim197^\circ$ is predicted for orbits with small radial distances when the surface is void of a central mass.  We then show that for small radii and large central masses, when the orbiting marble is deep within the well, the angular separation between successive apocenters transitions to values greater than 360$^\circ$.  We lastly compare these expressions to those describing elliptical-like orbits about a static, spherically-symmetric massive object in the presence of a constant vacuum energy, as described by general relativity.

\end{abstract}\maketitle

\section{Introduction}

Within the last several years there has been a wealth of papers published in this journal regarding the motion of an orbiting body on a cylindrically-symmetric surface, such as a funnel or a circular spandex fabric, residing in a uniform gravitational field.  Some of the interest in this topic can be attributed to fact that the orbital dynamics of an object on such a two-dimensional (2D) surface offers a non-trivial and interesting application of Lagrangian dynamics.  With the help of a camera capable of generating short video clips and one of the aforementioned surfaces, one can easily extend this theoretical study to the experimental domain.  This combined theoretical and experimental approach offers comprehensive undergraduate research possibilities as the relevant physics is accessible to any student who has taken an advanced dynamics course.\cite{weller} 

Perhaps some of the interest in this topic arises from the fact that an orbiting body on a 2D surface is often used as a conceptual analogy when describing particle orbits about massive objects in general relativity.  In Einstein's theory of general relativity (GR), gravity is described as the warping of space and time due to the presence of matter and energy.  Although in Newtonian theory gravity is understood as the force of one massive object on another, in GR gravity emerges as a physical phenomenon due to the fact that a massive object warps the spacetime around it.   According to GR, the planets move along geodesics, or free-particle orbits, in the non-Euclidean geometry about the sun.  Hence, the planets move in their elliptical-like orbits aroung the sun not due to a `force' per se but rather due to the fact that the spacetime around the sun is curved by its presence.

In a typical introductory GR course, the spacetime external to a static, spherically-symmetric massive object is often extensively studied as it offers one of the simplest, exact solutions to the field equations of GR.  When geodesics about this object are then encountered, the conceptual analogy of a marble rolling on a taut elastic fabric warped by a massive compact object, such as a bowling ball, is often embraced in the discussion.   When the rolling marble is far from the massive object, the elastic fabric is flat and the marble moves in a straight line at a constant speed in agreement with Newton's first law.  When the rolling marble approaches the massive object, as the analogy goes, it moves in a curved path not due to a force acting on the marble, but rather due to the fact that the space in which it resides (the 2D surface) is warped by the object. 
This analogy offers the student of GR a way of visualizing a warped spacetime and should not be confused with embedding diagrams.\cite{embedding}  However, with this being said, it has been shown that there exists no 2D cylindrically-symmetric surface residing in a uniform gravitational field that can generate the exact Newtonian orbits of planetary motion, except for the special case of circular orbits.\cite{English}  This null result was shown to hold when considering the fully general relativistic treatment of planetary orbits around non-rotating, spherically-symmetric  objects.\cite{me2} 

White and Walker first explored marbles orbiting on a warped spandex surface by applying Newtonian mechanics to a marble residing on a cylindrically-symmetric spandex fabric subjected to a central mass.  There they arrived at an expression describing the approximate shape of the spandex surface and found that a marble orbiting upon this surface obeys the relation $T^3\propto r^2$ for circular orbits in the small slope regime.  This fascinating result is highly reminiscent of Kepler's third law for planetary motion, but with the powers transposed.\cite{GW}  Lemons and Lipscombe then showed that the shape of the spandex surface can equivalently be found by minimizing the total potential energy of the spandex fabric through the use of the calculus of variations.  The total potential energy considered includes the elastic potential energy of the spandex surface and the gravitational potential energy of the central mass.  The calculus of variations approach proves advantageous over the Newtonian treatment as the results of White and Walker emerge as a special case of a more general treatment, where a pre-stretch of the fabric can be included in the analysis.\cite{DL}  This work was later expanded upon by Middleton and Langston, where the authors considered circular orbits on the spandex fabric in \textit{both} the small and large slope regimes.  There they generalized the aforementioned calculus of variations treatment by including the previously-neglected gravitational potential energy of the spandex surface to the total potential energy of the spandex/central mass system.  They showed that the mass of the spandex fabric interior to the orbital path influences the motion of the marble and can even dominate the orbital characteristics.  They also showed that the modulus of elasticity of the spandex fabric is not constant and is itself a function of the stretch.\cite{me2}  Later, Nauenberg considered precessing elliptical-like orbits with small eccentricities on cylindrically-symmetric funnels with various shape profiles.  By employing a perturbative method, an approximate solution to the orbital equation of motion was obtained where the precession parameter was found to be determined uniquely by the slope of the cylindrically-symmetric surface.\cite{Nauenberg}  This perturbative method was then used to find the 2D surfaces that are capable of generating the stationary elliptical orbits of Newtonian gravitation and the precessing elliptical-like orbits of GR, for orbits with small eccentricities.\cite{me}  Here, we're interested in understanding elliptical-like orbits of a marble on a warped spandex fabric.  

Our paper is outlined as follows.  In Sec. \ref{eqn}, we first consider the motion of a rolling marble on an arbitrary, cylindrically-symmetric surface and present the orbital equation of motion.  
For elliptical-like orbits with small eccentricities, we solve this equation of motion perturbatively and find a valid solution when the precession parameter obeys an algebraic relation involving the slope of the surface. We then present the equation that describes the slope of an elastic surface, which arises from minimizing the total potential energy of the spandex/central mass system.  In Sec. \ref{orbits} and Sec. \ref{orbits2}, we explore the elliptical-like orbits of the rolling marble in the small and large slope regimes, respectively.  We arrive at an expression describing the angular separation between successive like-apsides in each regime and compare with our experimental results.\cite{apocenter}  Lastly, in Appendix \ref{AdS}, we outline the general relativistic treatment of obtaining elliptical-like orbits with small eccentricities about a static, spherically-symmetric massive object in the presence of a vacuum energy.  After arriving at an expression describing the angular separation between successive like-apsides, we compare it to its counterpart for the marble on the spandex surface in both the small and large slope regimes. 

\section{Elliptical-like orbits of a marble rolling on a cylindrically-symmetric elastic surface} \label{eqn}
We begin this manuscript by presenting the equations of motion that describe a spherical marble of uniform mass density rolling on a cylindrically-symmetric surface in a uniform gravitational field.  These equations can be obtained by first constructing the Lagrangian that describes a rolling marble, which is constrained to reside upon a cylindrically-symmetric surface, in cylindrical coordinates $(r,\phi,z)$.  This Lagrangian includes both the translational and rotational kinetic energy of the rolling marble and the gravitational potential energy of the orbiting body. The resultant Lagrangian can then be subjected to the Euler-Lagrange equations, which yield the equations of motion of the form
\bea
(1+z'^2)\ddot{r}+z'z''\dot{r}^2-r\dot{\phi}^2+\frac{5}{7}gz'&=&0,\label{roft}\\
\dot{\phi}&=&\frac{5\ell}{7r^2},\label{phi}
\eea
where a dot indicates a time-derivative, $z'\equiv dz/dr$, and $\ell$ is the conserved angular momentum per unit mass.\cite{Marion} It is noted that these equations of motion have been derived in full detail elsewhere.\cite{me2, English}   In arriving at Eqs. (\ref{roft}) and (\ref{phi}), it is noted that the scalar approximation for an object that rolls without slipping was employed.  A full vector treatment of the intrinsic angular momentum of a rolling object on a cylindrically-symmetric surface can be found elsewhere.\cite{white}

Notice that one can easily decouple the above differential equations by inserting Eq. (\ref{phi}) into Eq. (\ref{roft}).  The resulting differential equation can then be solved for $r(t)$, at least in principle, once the equation describing the cylindrically-symmetric surface, $z(r)$, is specified.  Here we are interested in arriving at an expression for the radial distance in terms of the azimuthal angle, $r(\phi)$.  Using the chain rule and Eq. (\ref{phi}), we construct a differential operator of the form
\be\label{oper}
\frac{d}{dt}=\frac{5\ell}{7r^2}\frac{d}{d\phi}.
\ee
By employing Eqs. (\ref{phi}) and (\ref{oper}), Eq. (\ref{roft}) can be transformed into an orbital equation of motion of the form
\be\label{rofphi}
(1+z'^2)\frac{d^2r}{d\phi^2}+(z'z''-\frac{2}{r}(1+z'^2))\left(\frac{dr}{d\phi}\right)^2-r+\frac{7g}{5\ell^2}\cdot z'r^4=0.
\ee
This dynamical equation of motion describes the radial distance of the orbiting marble in terms of the azimuthal angle, for a given cylindrically-symmetric surface specified by $z(r)$, and equates to a non-linear differential equation that can be solved perturbatively.  As here we are interested in studying elliptical-like orbits, we choose an approximate solution for the radial distance of the orbiting body to be of the form
\be\label{sol}
r(\phi)=r_0(1-\varepsilon\cos(\nu\phi)),
\ee
where $\varepsilon$ is the eccentricity of the orbit and $\nu$ is the precession parameter.  For an orbiting body whose radial distance is described by Eq. (\ref{sol}), $r_0$ represents the average radial distance of the orbital motion.  We note that in what follows, $\varepsilon$ will be treated as small and used as our expansion parameter, whereas the parameters $\nu$ and $r_0$ will be determined by Eq. (\ref{rofphi}).

The precession parameter, $\nu$, can be further understood through the relation 
\be\label{delphi}
\nu\equiv\frac{360^\circ}{\Delta\phi},
\ee
where $\Delta\phi$ represents the angular separation between two like-apsides (the angle between two consecutive apocenters or two consecutive pericenters) of the orbital motion.\cite{apocenter}  Notice that for $\nu=1$, Eq. (\ref{sol}) equates to the perturbative approximation of the \textit{exact} solution of Newtonian gravitation, whose orbits equate to the well-known conic sections, to first-order in the eccentricity.\cite{conic}  When the azimuthal angle advances by $360^\circ$, the orbiting body will find itself at the same radial distance and will precisely retrace its preceding orbital path.  Hence, for $\nu=1$, Eq. (\ref{sol}) approximately describes closed or stationary elliptical orbits with small eccentricities.  When $\nu\neq 1$, $\Delta\phi\neq 360^\circ$ and Eq. (\ref{sol}) approximately describes precessing elliptical-like orbits with small eccentricities.  Notice that for $\nu>1$, Eq. (\ref{delphi}) implies that the angular separation between two consecutive like-apsides will be \textit{less than} $360^\circ$ and the apsides will `march backwards' in the azimuthal direction over successive orbits.  Contrarily, for $\nu<1$, the angular separation between two consecutive like apsides will be \textit{greater than} $360^\circ$ and the apsides will `march forward' in the azimuthal direction.

Inserting Eq. (\ref{sol}) into Eq. (\ref{rofphi}) and keeping terms up to first-order in $\varepsilon$, we find a valid approximate solution to the orbital equation of motion when the algebraic expressions
\bea
\ell^2&=&\frac{7}{5}gr_0^3z_0'\label{ell}\\
\nu^2&=&\frac{3z_0'+r_0z_0''}{z_0'(1+z_0'^2)}\label{nu}
\eea
are satisfied, where $z_0'$ and $z_0''$ are radial derivatives of $z$ evaluated at the average radial distance, $r_0$.   In arriving at the above expressions, it is noted that the slope and the radial derivative of the slope were expanded about the average radial distance of the respective orbit.  The above expressions were presented in a previous work\cite{Nauenberg} and then later used to find the 2D surfaces that generate Newtonian and general relativistic orbits with small eccentricities.\cite{me}  Equation (\ref{ell}) specifies the necessary angular momentum per unit mass needed, at a given average radius and for a given slope of the surface, for circular or elliptical-like orbits to occur.  Equation (\ref{nu}) yields a prediction for the precession parameter in terms of the slope and the radial derivative of the slope of a given surface at a given average radial distance.  

In the subsection that follows we present the equation that describes the slope of a cylindrically-symmetric spandex fabric, subjected to a central mass placed upon its surface, in a uniform gravitational field.   This differential equation yields approximate expressions for the slope in the small and large slope regimes, which can be inserted into Eq. (\ref{nu}) to yield a theoretical prediction for the precession parameter in each regime.  Using Eq. (\ref{delphi}), the approximate solution for the angular separation between two consecutive like-apsides for a marble orbiting on this spandex surface can be found in both the small and large slope regimes. 

\subsection{The slope of an elastic fabric warped by a central mass}\label{Shape}
The equation that describes the slope of a cylindrically-symmetric spandex surface residing in a uniform gravitational field can be obtained through the use of the calculus of variations, as the surface will take the shape that \textit{minimizes} the total potential energy of the central mass/elastic fabric system.  The total potential energy can be written as an integral functional, where we take into account the elastic potential energy of the spandex fabric and the gravitational potential energy of the surface of uniform areal mass density $\sigma_0$ and of the central object of mass $M$.  This resultant functional can then be subjected to Euler's equation, which yields a differential equation that describes the slope of the spandex fabric.\cite{Marion}
This resultant differential equation can be integrated and yields the expression
\be\label{shape}
rz'\left(1-\frac{1}{\sqrt{1+z'^2}}\right)=\alpha\left(M+\sigma_0\pi r^2\right),
\ee
where we defined the parameter $\alpha\equiv g/2\pi E$, where $E$ is the modulus of elasticity of the elastic material.  We note that this differential equation that describes the slope of the elastic surface was previously derived in an earlier work of one of the authors.\cite{me2}

Notice that Eq. (\ref{shape}) describes the slope of the spandex fabric at radial distances $r<R$, where $R$ is the radius of the spandex surface, and is dependent upon the modulus of elasticity, the mass of the central object, and the mass of the spandex fabric interior to this radial distance, which is explicitly seen through the $\sigma_0\pi r^2$ term.  Equation (\ref{shape}) corresponds to a quartic equation, which can be solved exactly for the slope, $z'(r)$, but cannot be integrated analytically to yield the shape of the surface, $z(r)$.  The exact expression for $z'$ is rather awkward to work with but can be conveniently approximated in the small and large slope regimes.  These regimes correspond to where the surface is nearly horizontal and nearly vertical, respectively.  

The theoretical work outlined in this section offers an interesting and non-trivial application of Lagrangian dynamics and of the calculus of variations for an advanced undergraduate student.  For the student who wishes to forego these derivations and focus on the experimental aspects of this project, Eqs. \eqref{roft}, \eqref{phi}, and \eqref{shape} can easily serve as a starting point.

In the next section we explore the behavior of the slope of the spandex surface in the small slope regime and, using Eq. (\ref{nu}), arrive at an expression for the precession parameter.  Interestingly, this theoretical prediction for the precession parameter can then easily be compared to experiment by directly measuring the angular separation between successive like-apsides and then comparing to the theoretical prediction of Eq. (\ref{delphi}).

\section{Elliptical-like orbits in the small slope regime}\label{orbits}
We first explore the behavior of the angular separation between consecutive like-apsides in the regime where the slope of the spandex surface is small and the surface nearly horizontal.  This physically corresponds to the region near the perimeter of the spandex fabric when a small central mass (or no central mass) has been placed upon the surface.  In this small slope regime where $z'\ll 1$, we can expand the square root in the denominator of Eq. (\ref{shape}) in powers of $z'^2$.  Keeping only the lowest-order non-vanishing contribution and then solving Eq. (\ref{shape}) for $z'$, we obtain an approximate expression describing the slope of the elastic surface of the form
\be\label{smallslope}
z'(r)\simeq\left(\frac{2\alpha}{r}\right)^{1/3}\left(M+\sigma_0\pi r^2\right)^{1/3}.
\ee
Inserting Eq. (\ref{smallslope}) into Eq. (\ref{nu}) yields an approximate expression for the precession parameter when the slope of the surface is evaluated at the average radial distance, $r_0$, of the orbiting body.  Plugging this expression for $\nu$ into Eq. (\ref{delphi}) and solving for $\Delta\phi$, we obtain an expression predicting the value of the angular separation between consecutive apocenters (or between consecutive pericenters), which takes the form
\be\label{smallsol}
\Delta{\phi}=220^\circ\left[1+\left(2\alpha(M+\sigma_0\pi r_0^2)/r_0\right)^{2/3}\right]^{1/2}\left[1+\frac{1}{4}\frac{\sigma_0\pi r_0^2}{\left(M+\sigma_0\pi r_0^2\right)}\right]^{-1/2}.
\ee

Upon inspection of Eq. (\ref{smallsol}), it is noted that the value of this angular separation between successive like-apsides is dependent upon the mass of the central object, $M$, and the mass of the spandex fabric interior to the average radial distance of the elliptical-like orbit, $\sigma_0\pi r_0^2$.  These two competing terms are on equal footing when $M\simeq\sigma_0\pi r_0^2$.  For our spandex fabric, we measured the uniform areal mass density to be $\sigma_0=0.172$ kg/m$^2$ and found that the average radial distance of the elliptical-like orbits in the small slope regime is $r_0\simeq 0.33$ m.  Using these measured values, we find that these two terms are on equal footing for a central mass of about $M\simeq 0.059$ kg.  Hence, when the mass of the central object is less than $0.059$ kg, the mass of the spandex fabric interior to the average radial distance of the elliptical-like orbit becomes the dominant factor in determining the angular separation between consecutive like-apsides, as described by Eq. (\ref{smallsol}).  When the mass of the central object is much smaller than the mass term of the spandex fabric, Eq. (\ref{smallsol}) yields a limiting behavior for the angular separation of the form
\be\label{smallM}
\lim_{M\ll\sigma_0\pi r_0^2}\Delta{\phi}\simeq 197^\circ\left[1+(2\alpha\sigma_0\pi r_0)^{2/3}\right]^{1/2}.
\ee
Conversely, when the mass of the central object dominates over the mass term of the spandex fabric, Eq. (\ref{smallsol}) yields a limiting behavior for the angular separation of the form
\be\label{smallsigma}
\lim_{M\gg\sigma_0\pi r_0^2}\Delta{\phi}\simeq 220^\circ\left[1+\left(\frac{2\alpha M}{r_0}\right)^{2/3}\right]^{1/2},
\ee
where, physically, one must be careful to ensure that the elliptical-like orbits are still occurring in the small slope regime with this relatively large central mass.  

Upon comparison of the above approximate expressions, notice that Eq. (\ref{smallM}) predicts that the angular separation increases with increasing average radial distance whereas Eq. (\ref{smallsigma}) predicts a decreasing angular separation with increasing average radial distance, for a given central mass.  Further notice that Eq. (\ref{smallsol}) predicts a minimal angular separation of $\Delta\phi_{min}\simeq 197^\circ$, which occurs for orbits with vanishingly small average radial distances when the surface is void of a central mass.  


\subsection{The experiment in the small slope regime}
To experimentally test our theoretical prediction for the angular separation between successive apocenters and pericenters in the small slope regime, we needed a cylindrically-symmetric spandex fabric, capable of supporting elliptical-like orbits of a marble upon its surface, and a means of precisely determining the polar coordinates of the orbiting marble.  A 4-ft diameter mini-trampoline was stripped of its trampoline surface and springs.  A styrofoam insert was fashioned to exactly fit the interior of the trampoline frame, which collectively formed a flat, table-like structure.  With the styrofoam insert in place, a nylon-spandex fabric was draped over the trampoline frame, where we ensured that no pre-stretch or slack was introduced.  The fabric was then securely fastened to the frame by means of a ratchet strap and the styrofoam insert was removed, allowing the fabric to hang freely.

\begin{figure}[!t]
\centering
\includegraphics[width=15cm]{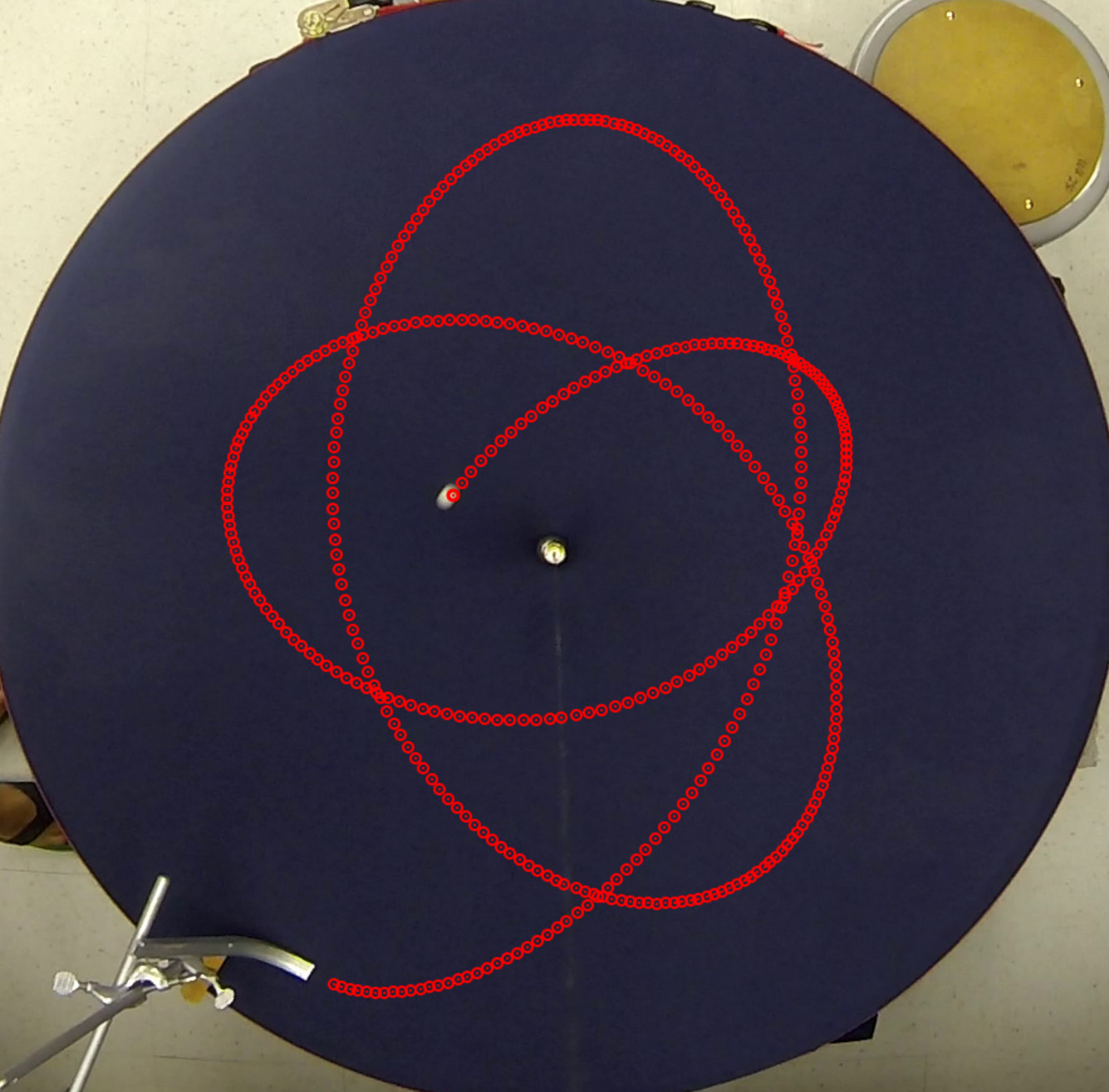}
\caption{Screenshot of elliptical-like orbits of a marble rolling on the spandex surface, here subjected to a central mass of $M=0.199$ kg, in the small slope regime.  Each circle marks the position of the rolling marble, in intervals of 1/60 of a second, as it orbits the central mass.  The three orbits imaged here were found to have angular separations between successive apocenters of $\Delta\phi=213.5^\circ, \;225.7^\circ,\; 223.9^\circ$ and eccentricities of $\varepsilon=0.31, \;0.29, \;0.31$, respectively.}
\label{fig:smallorbit}
\end{figure}

It is noted that the shape of the freely hanging spandex fabric can easily be obtained.  When the spandex fabric is void of a central mass the slope of the surface is small, hence, the approximation for the slope given by Eq. (\ref{smallslope}) is valid everywhere on the surface.  Setting $M=0$ in Eq. (\ref{smallslope}) and integrating we obtain the height of the surface, which is given by the expression
\be\label{shape0}
z(r)_{M=0}=\frac{3}{4}(2\alpha\sigma_0\pi)^{1/3}(r^{4/3}-R^{4/3}),
\ee
where $R=0.585$ m is the inner radius of the trampoline frame and we chose the zero point of our coordinate system to coincide with $z(R)=0$.  By measuring the height of the spandex fabric at the center of the apparatus, which was found to be $z(0)=-0.098$ m, and by knowing the areal mass density of the spandex fabric, we experimentally obtain a value of $\alpha=0.018$ m/kg.  Having an experimental value for $\alpha$ is necessary as we wish to compare our predicted values for the angular separation between any successive like-apsides to the experimentally measured values. 

A camera capable of taking video clips at a rate of 60 frames per second was mounted above the spandex surface.  A plumb line, strung from the camera's perched location, was used to align the trampoline frame below and to determine the precise center of the fabric where the central mass should be placed.  A ramp was mounted near the perimeter of the trampoline and orientated in the tangential direction so elliptical-like orbits with varying eccentricities, and hence varying average radial distances, could be generated by simply varying the initial speed of the marble.  Several short video clips of elliptical-like orbits were then generated for a range of central masses and eccentricities.  The video clips were imported into Tracker,\cite{Tracker} a video-analysis and modeling software program, where the marble's radial and angular coordinates could readily be extracted one frame at a time (see Fig. \ref{fig:smallorbit}).  The apocenters and pericenters of the orbiting marble were identified by scrolling through the instantaneous radii and determining the maximum and minimum values, along with the accompanying angular coordinates. To remain within the small slope regime, only the first five orbits of a given run were considered, which yielded a maximum of ten data points per run.  These data were then imported into Excel for analysis.

\begin{figure}[!t]
\centering
\includegraphics[width=15cm]{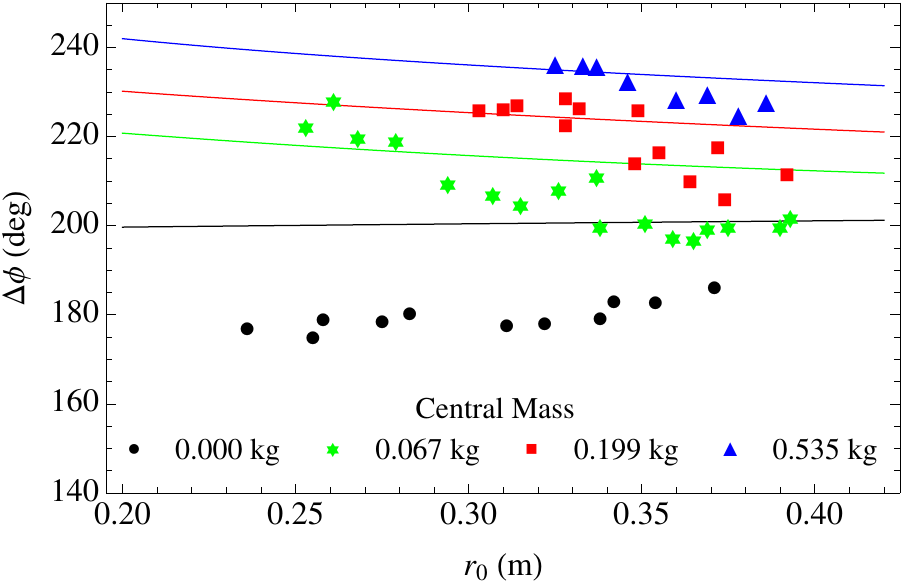}
\caption{Plot of the angular separation between two successive like-apsides, $\Delta\phi$, versus the average radial distance, $r_0$, of elliptical-like orbits in the small slope regime for central masses ranging from $M=0$ kg to $M=0.535$ kg. } 
\label{fig:SmallSlope}
\end{figure}

The angular separation between successive like-apsides could easily be obtained from the measured angular coordinates.  The average radial distance, $r_0$, was found by averaging over the measured instantaneous radii over one orbit; one data point was hence generated from these two quantities.  It is noted that the radius of a successive apocenter or pericenter was found to decrease (dramatically in the small slope regime) due to frictional losses and drag.  To account for this loss, the average radial distance of two successive apocenters or pericenters was calculated and used in the analysis in determining the eccentricity of an orbit.\cite{eccentricity}  Elliptical-like orbits with eccentricities falling within the range of $0.15\leq\varepsilon<0.35$ were kept for analysis whereas orbits with eccentricities outside of this adopted range were systematically discarded.

Using our aforementioned measured value of $\alpha$, the theoretical curves of the angular separation between successive like-apsides, as described by Eq. (\ref{smallsol}), can be plotted versus the average radial distances, $r_0$, of an elliptical-like orbit for a given central mass.  These theoretical curves can then be compared to the measured values of $\Delta\phi$, both are collectively displayed in Fig. \ref{fig:SmallSlope}.  

Upon examination of Fig. \ref{fig:SmallSlope}, notice that the measured values of $\Delta\phi$ tend to reside below the theoretical curves, with larger discrepancies occurring for smaller central masses.  The authors hypothesize that this is predominantly due to an increased frictional loss for an orbiting marble about smaller central masses.  As the central mass decreases, the stretching of the spandex fabric decreases, and the warping of the spandex fabric due to the orbiting marble's weight increases.  This generates a scenario were the marble is having to effectively climb out of its own gravitational well, more so then for a larger central mass, and encounters an increased frictional loss resulting in the marble reaching its apocenter earlier than it would without this loss.  This can be compared to 2D projectile motion where the projectile is subjected to a retarding force.  For a projectile subjected to a large retarding force, one finds that it reaches a smaller peak height in a shorter horizontal distance than it would for a smaller retarding force.  Further note that the measured values for $\Delta\phi$ tend to increase for increasing average radial distance for zero central mass and decrease for increasing average radial distances for non-zero central mass, inline with the theoretical predictions.  

In the next section, we explore the behavior of the angular separation between successive apocenters in the large slope regime.  We find that within this regime,  the angular separation takes on values \textit{greater than} 360$^\circ$ for very small radial distances and large central masses.  This corresponds to the apsides marching forward in the azimuthal direction over successive orbits.  Interestingly, elliptical-like orbits around non-rotating, spherically-symmetric massive objects in general relativity are also found to have angular separations between successive like-apsides taking on values greater than 360$^\circ$.  A comparison between these GR orbits and the orbits on a spandex fabric will be discussed in the appendix.

\section{Elliptical-like orbits in the large slope regime}\label{orbits2}
We now wish to explore the behavior of the angular separation between successive apocenters in the large slope regime.  This regime corresponds physically to the region deep inside the well, namely, on the spandex fabric where $r_0$ is small, when a large central mass has been placed upon the surface.  In this large slope regime, where $z'\gg 1$, we can again expand the square root in the denominator of Eq. (\ref{shape}) by first pulling $z'$ out of the square root and then expanding in powers of $1/z'^2$.  Keeping only the lowest-order contribution in this expression and then solving Eq. (\ref{shape}) for $z'$, the slope takes on the approximate form
\be\label{largeslope}
z'(r)\simeq 1+\frac{\alpha M}{r}, 
\ee
where we dropped the areal mass density term as it is insignificant in this regime.  Now inserting Eq. (\ref{largeslope}) into Eq. (\ref{nu}) and then plugging this expression for $\nu$ into Eq. (\ref{delphi}),  we arrive at an expression describing the angular separation between successive apocenters in the large slope regime, which takes the form
\be\label{largesol}
\Delta\phi=360^\circ\sqrt{\frac{(1+\alpha M/r_0)}{(3+2\alpha M/r_0)}\left[1+\left(1+\frac{\alpha M}{r_0}\right)^2\right]},
\ee
where we evaluated $r$ at the average radial distance, $r_0$, of the elliptical-like orbits. 
Notice that this expression, like its counterpart in the small slope regime, also has two terms competing for dominance.  However, as the above expression is only valid for $z'\gg 1$, with $z'$ approximately given by Eq. (\ref{largeslope}), Eq. (\ref{largesol}) isn't valid when $\alpha M/r_0\ll 1$.  

For a large central mass and a \textit{very} small average radial distance, when the orbiting marble is deep within the well, Eq. (\ref{largesol}) yields a limiting behavior for the angular separation of the form
\be\label{largesolapprox}
\lim_{\alpha M/r_0\gg 1}\Delta\phi\simeq\frac{360^\circ}{\sqrt{2}}\cdot\frac{\alpha M}{r_0}.
\ee
Notice that when $\alpha M/r_0>\sqrt{2}$, Eq. (\ref{largesolapprox}) predicts that $\Delta\phi>360^\circ$.  Hence, as $\alpha M/r_0$ approaches and then becomes larger than the $\sqrt{2}$, successive apocenters are predicted to transition from marching backwards to marching forwards in the azimuthal plane over successive orbits.

\subsection{The experiment in the large slope regime}
We now wish to compare our theoretical prediction for the angular separation in the large slope regime, given by Eq. (\ref{largesol}), to our experimentally measured values.\cite{apsides} If the modulus of elasticity was truly constant for the spandex fabric, then the experimentally obtained value of $\alpha=0.018$ m/kg in the small slope regime could be used here.  This, however, is not the case and has been discussed extensively elsewhere.\cite{me2}  Although the modulus of elasticity has been shown to be a function of the stretch, it is approximately constant in each regime.  Therefore, the analysis presented in this manuscript is valid when applied to each regime separately.  
\begin{figure}[!t]
\centering
\includegraphics[width=15cm]{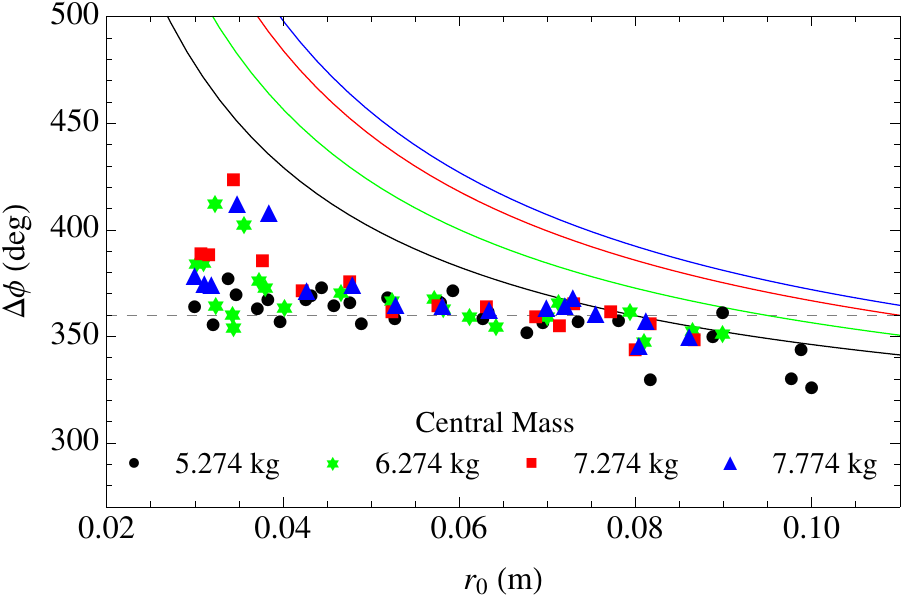}
\caption{Plots of the angular separation between two successive apocenters, $\Delta\phi$, versus the average radial distance, $r_0$, of elliptical-like orbits in the large slope regime for central masses that range from $M=5.274$ kg to $M=7.774$ kg. }
\label{fig:LargeSlope}
\end{figure}

To obtain $\alpha$ in this large slope regime, we note the slope of the spandex surface, given by Eq. (\ref{largeslope}), can be integrated, which yields the height of the spandex fabric as a function of the radius and the central mass.  Performing this integration and then evaluating at a small fixed radius $R_B=0.0154$ m on the spandex fabric, we obtain the expression
\be\label{alphaLS}
\frac{z(M)}{\ln(R_B)}=\alpha (M-M_0), 
\ee
where, here, we set $z(M_0)=0$ where $M_0$ corresponds to the smallest central mass probed.  This expression gives a scaled height of the fabric as a function of the central mass, $M$, for a given radius of the fabric, $R_B$, in the large slope regime.  As all of the quantities in Eq. (\ref{alphaLS}) save $\alpha$ can be  measured directly, the slope of the best fit line of a plot $z(M)/\ln(R_B)$ vs $(M-M_0)$ yields the value of the parameter $\alpha$.  By orientating our camera to obtain a side view of the fabric's surface, we subjected our spandex fabric to the same range of central masses used in generating particle orbits in this large slope regime.  We then imported this video of the shape profile of the spandex fabric into Tracker and measured the heights of the fabric at the aforementioned fixed radius as a function of the central mass $M$.  
The slope of this best fit line yields a value of $\alpha=0.0049$ m/kg.


The experimental setup and procedure for probing the large slope regime of the spandex fabric was quite similar to that of the small slope regime, with a few notable exceptions.  As here we wish to probe small radial distances for large central masses, we positioned a small spherical object of radius $\sim1.5$ cm in the center of the fabric.  A sling, capable of supporting several kilograms of mass, was fashioned and fastened to the spherical object from below the fabric with the help of an adjustable pipe clamp.  This setup allowed us to probe a range of large central masses while maintaining a constant diameter circular well.

The average radii of successive orbits do not decrease as drastically when the orbiting body is deep within the well as compared to orbits in the small slope regime, hence many orbits can potentially be analyzed for any given run.  The angular separations, $\Delta\phi$, were found and the corresponding average radial distances, $r_0$, were obtained.  Again, ellliptical-like orbits with eccentricities found to lie within the range $0.15\leq\varepsilon<0.35$ were kept and further analyzed, those not in this defined range were systematically discarded.

Using our experimentally determined value of $\alpha=0.0049$ m/kg, the theoretical curves of the angular separation, $\Delta\phi$, described by Eq. (\ref{largesol}), can be plotted versus the average radial distance, $r_0$, of an elliptical-like orbit for a given central mass along with the measured values and are collectively displayed in Fig. \ref{fig:LargeSlope}.  Notice that as the average radial distance decreases, the angular separation increases.  The horizontal dashed line displayed in Fig. \ref{fig:LargeSlope} corresponds to $\Delta\phi=360^\circ$.  Hence, for data points lying above this line, the apocenter marched forward in the azimuthal direction when compared to that of the previous orbits. 

We note that when the average radial distances of the elliptical-like orbits were greater than $\sim10$ cm, the measured values of $\Delta\phi$ were found to become substantially smaller than those shown in Fig. \ref{fig:LargeSlope}.  As our theoretical prediction for $\Delta\phi$ is only valid in the large slope regime, we neglected the data points beyond this radial distance.  For radial distances smaller than $\sim3$ cm, the measured values of $\Delta\phi$ were found to be quite sporadic.  For orbits with small average radial distances, the angular velocity becomes quite large and hence only a handful of particle locations are imaged ($\sim12$) for a given orbit; this yields an angular separation of $\sim30^\circ$ between imaged locations and gives rise to a fairly large uncertainty in the angular measurement.  Here we display the data points with average radial distances $3.0\;\mbox{cm}\leq r_0\leq 10\;\mbox{cm}$.

\section{conclusion}
In this manuscript, we arrived at expressions describing the angular separation between successive like-apsides for a marble rolling on a cylindrically-symmetric spandex surface in a uniform gravitational field in both the small and large slope regimes. We found that a minimal angular separation of $\sim197^\circ$ was predicted for orbits with small radial distances when the surface is void of a central mass.  We also showed that for a marble orbiting with a small radius about a large central mass, corresponding to when the marble is deep within the well, the angular separation between successive like-apsides takes on values greater than 360$^\circ$.  For these orbits, successive apocenters `march forward' in the azimuthal direction in the orbital plane, reminiscent of particle orbits about spherically-symmetric massive objects as described by general relativity.  These theoretical predictions were shown to be consistent with the experimental measurements.

Although the orbital characteristics of a marble on a warped spandex fabric fundamentally differ from those of particle orbits about a spherically-symmetric massive object as described by general relativity, the authors contend that embracing the conceptual analogy between the two is insightful for the student of GR.  Hence, this proposed theoretical/experimental undergraduate research project can be extended into the realm of general relativity, where a study of the elliptical-like orbits about a spherically-symmetric massive object with small eccentricities can be compared and contrasted to the theoretical work contained within this manuscript.

In the appendix, we study the physics of the analogy and find that some striking similarities exist between these elliptical-like orbits.  Although differing in functional form, we show that the angular separation between like-apsides for marbles orbiting in the large slope regime and for particles orbiting near the innermost stable circular orbit both diverge for vanishingly small radial distances in a somewhat similar manner.  We also show that the areal mass density of the spandex fabric plays the role of the vacuum energy density in GR, in regards to their competing relationship with their respective central mass, in the expressions for the angular separation.

\begin{appendix}

\section{Elliptical-like orbits with small eccentricities in general relativity}\label{AdS}
In this section, we outline the general relativistic treatment of obtaining elliptical-like orbits with small eccentricities about a spherically-symmetric massive object in the presence of a constant vacuum energy density, or equivalently, a cosmological constant.  We arrive at the necessary conditions for elliptical-like orbits to occur in this spacetime and compare and contrast the expression for the angular separation between consecutive like-apsides to that obtained for a rolling marble on the warped spandex fabric in \textit{both} the small and large slope regimes.  Although these expressions differ in their functional forms, there are some striking similarities.  Here we argue that a thorough study of the physics of the analogy, a marble orbiting on a spandex surface, can aid in the learning of free particle orbits, or geodesics, in general relativity.

The addition of the cosmological constant to the field equations of general relativity can be traced back to Einstein himself.  One of his first applications of this new theory of gravitation, which describes gravity as the warping of space and time due to the presence of matter and energy, was to the field of cosmology.  Einstein added the cosmological constant to the relativistic equations only after realizing that these equations fail to offer a static cosmological solution to describe the seemingly static universe of his time.  When the universe was later discovered to be expanding, the cosmological constant became an unnecessary addition and Einstein missed the opportunity of predicting an expanding universe, as his cosmological constant-free equations seemed to be implying.  With the much later discovery of a late-time \textit{accelerated} expansion of the universe, the importance of a cosmological constant has once again arisen, but with the cosmological constant now recast into the form of a vacuum energy density.  Mathematically, a cosmological constant, $\Lambda$, which finds itself on the curvature side of the field equations of GR, can be equivalently reformulated into a constant vacuum energy density, $\rho_0$, which resides on the matter/energy side.  These quantities are related through the algebraic relation
\be\label{rholambda}
\rho_0=(c^2/8\pi G)\Lambda, 
\ee
where $G$ is Newton's universal constant and $c$ is the speed of light.\cite{hartle,carroll}  The reason for the relabeling of the cosmological constant emerges in quantum field theory, where zero-point fluctuations, corresponding to particles and anti-particles coming into and out of existence, are predicted.  These vacuum fluctuations give rise to a nonzero vacuum energy density.  It is noted that the cosmological constant, and therefore the constant vacuum energy density, can be of either sign.  Observational evidence, however, suggests that we live in a universe with a small but nonzero, \textit{positive} vacuum energy density 
whose presence is needed to explain the observed accelerated expansion of the universe.  

Using the Schwarzschild coordinates $(t,r,\theta,\phi)$, the equations of motion describing an object of mass $m$ orbiting about a non-rotating, spherically symmetric massive object of mass $M$, in the presence of a cosmological constant, in general relativity are of the form
\bea
\ddot{r}+\frac{GM}{r^2}-\frac{\ell^2}{r^3}+\frac{3GM\ell^2}{c^2r^4}-\frac{1}{3}\Lambda c^2r&=&0,\label{releq}\\
\dot{\phi}&=&\frac{\ell}{r^2}\label{relphi},
\eea
where in this section a dot indicates a derivative with respect to proper time and $\ell$ is the conserved angular momentum per unit mass.\cite{Stuchlik,Hackmann,Cruz}  

It is noted that the last two terms in Eq. (\ref{releq}) correspond to relativistic terms which are not present in Newtonian gravitation.  By setting these two relativistic terms equal to zero, one obtains the equations of motion of Newtonian theory; these yield a solution for the radial distance that describes orbits that equate to the conic sections.\cite{conic}  The $1/r^4$ term equates to a general relativistic correction term to Newtonian gravitation, present even for a vanishing cosmological constant,  and offers a small-$r$ modification to Newtonian theory.  This correction term demands that the closed elliptical orbits of Newtonian theory are replaced with precessing elliptical-like orbits, which will become apparent in the paragraphs that follow.  For a nonzero cosmological constant, notice that the last term in Eq. (\ref{releq}) offers a large-$r$ modification to the orbits.

As in our treatment of elliptical-like orbits on a warped spandex surface, we are interested in obtaining the solution for the radial distance of the orbiting body in terms of the azimuthal angle.  Using the chain rule and Eq. (\ref{relphi}), we construct a differential operator of the form
\be\label{relop}
\frac{d}{d\tau}=\frac{\ell}{r^2}\frac{d}{d\phi}, 
\ee
where $\tau$ is the proper time.  By employing Eqs. (\ref{relphi}) and (\ref{relop}),  Eq. (\ref{releq}) can be transformed into an orbital equation of motion of the form
\be\label{releqphi}
\frac{d^2r}{d\phi^2}-\frac{2}{r}\left(\frac{dr}{d\phi}\right)^2+\frac{GM}{\ell^2}r^2-r+\frac{3GM}{c^2}-\frac{\Lambda c^2}{3\ell^2}r^5=0.
\ee
The solution to this orbital equation of motion describes the possible particle orbits about a non-rotating, spherically-symmetric massive object in the presence of a constant vacuum energy density.  

\subsection{$\Lambda=0$}
We wish to examine elliptical-like orbits with small eccentricities and first examine the case of a vanishing cosmological constant.  Inserting the perturbative solution of the form
\be\label{relsol}
r(\phi)=r_0(1-\varepsilon\cos(\nu\phi))
\ee
into Eq. (\ref{releqphi}), setting $\Lambda=0$, and keeping terms up to first-order in $\varepsilon$, we find a valid solution when the constants and parameters obey the relations
\bea
\ell^2&=&GMr_0\left(1-\frac{3GM}{c^2r_0}\right)^{-1},\label{relell2norho}\\
\nu^2&=&1-\frac{6GM}{c^2r_0}\label{relnu2norho}.
\eea
These algebraic expressions correspond to the conditions necessary for elliptical-like orbits with small eccentricities about a non-rotating, spherically-symmetric massive object, void of vacuum energy, and have been discussed elsewhere.\cite{me}  It is noted that when the general relativistic term $GM/c^2r_0$ is set equal to zero, we recover the stationary elliptical orbits of Newtonian gravitation, characterized by $\nu=1$.  Notice that $\nu$ becomes complex when $r_0<r_{ISCO}\equiv 6GM/c^2$; this implies that elliptical-like orbits are \textit{only} allowed for radii larger than this limiting value.  This threshold radius, $r_{ISCO}$, is known as the innermost stable circular orbit.  Further notice that for radii smaller than $r_0<3GM/c^2$, both the angular momentum per unit mass, $\ell$, and the parameter $\nu$ become complex.  This implies that there are no circular orbits, stable or unstable, allowed for an object orbiting about a non-rotating, spherically-symmetric massive object for these small radii. 

Using Eqs. (\ref{delphi}) and (\ref{relnu2norho}), the angular separation between successive apocenters takes the form
\be\label{reldelphinorho}
\Delta\phi=360^\circ\left(1-\frac{ r_{ISCO}}{r_0}\right)^{-1/2},
\ee
where we used our aforementioned definition of $r_{ISCO}$.  Note that as the average radius, $r_0$, of an elliptical-like orbit with a small eccentricity takes on values that approach the innermost stable circular orbit, $r_{ISCO}$, the angular separation between successive apocenters increases without bound.  To further explore this limiting behavior, we define
\be\label{smallr}
r_0\equiv r_{ISCO}+r,
\ee
where $r\ll r_{ISCO}$ is a vanishingly small radial distance.  Inserting Eq. (\ref{smallr}) into Eq. (\ref{reldelphinorho}) and expanding for small $r$, we find the limiting behavior for the angular separation to be of the form
\be\label{rll6GM}
\lim_{r\ll 6GM/c^2}\Delta\phi\simeq 360^\circ\sqrt{6}\cdot\sqrt{\frac{GM/c^2}{r}},
\ee
near the innermost stable circular orbit.  Recall that the functional form of the angular separation for a marble orbiting on the warped spandex fabric in the large slope regime took on a limiting behavior of the form
\be\label{rllaM}
\lim_{r_0\ll \alpha M}\Delta\phi\simeq\frac{360^\circ}{\sqrt{2}}\cdot\frac{\alpha M}{r_0}.
\ee
Upon comparison of Eqs. (\ref{rll6GM}) and (\ref{rllaM}), we note that both expressions diverge in the limit of vanishingly small distances $r_0$ and $r$, however, their functional forms differ. $\Delta\phi\sim 1/\sqrt{r}$ for the orbiting object about the non-rotating, spherically-symmetric massive object near the innermost stable circular orbit whereas $\Delta\phi\sim 1/r_0$ for the rolling marble on the spandex fabric in the large slope regime for small radial distances.  In the vanishingly small $r_0$ regime the slope of the spandex surface diverges, which can be seen upon inspection of Eq. (\ref{largeslope}), when the central mass is assumed to be point-like.  We further note that the parameter $\alpha\equiv g/2\pi E$ plays the role of $G/c^2$ in the particle orbits of general relativity.  Both constants set the scale of their respective theories and determine the value of the angular separation for a given central mass and average radial distance.

\subsection{$\Lambda\neq 0$}
We wish to push the conceptual analogy further between particle orbits in general relativity and rolling marble orbits on a warped spandex fabric in the small slope regime by considering the functional form of the angular separation for a nonzero cosmological constant.  Inserting Eq. (\ref{relsol}) into Eq. (\ref{releqphi}), with $\Lambda\neq 0$, and again keeping terms up to first-order in $\varepsilon$, we find that Eq. (\ref{relsol}) is a valid approximate solution when 
\bea
\ell^2&=&Gr_0\left(1-\frac{3GM}{c^2r_0}\right)^{-1}(M-2\cdot\frac{4}{3}\pi r_0^3\cdot\rho_0),\label{relell2}\\
\nu^2&=&1-\frac{6GM}{c^2r_0}-6\left(1-\frac{3GM}{c^2r_0}\right)\frac{ \frac{4}{3}\pi r_0^3\cdot\rho_0}{(M-2\cdot \frac{4}{3}\pi r_0^3\cdot\rho_0)}\label{relnu2},
\eea
where we eliminated $\Lambda$ in favor of $\rho_0$ by using Eq. (\ref{rholambda}). 
Now inserting Eq. (\ref{relnu2}) into Eq. (\ref{delphi}) and solving for $\Delta\phi$, we obtain the relativistic expression for the angular separation between consecutive like-apsides, which takes the form
\be\label{reldelphi}
\Delta\phi=360^\circ\left(1-\frac{6GM}{c^2r_0}\right)^{-1/2}\left[1-6\left(\frac{1-3GM/c^2r_0}{1-6GM/c^2r_0}\right)\frac{\frac{4}{3}\pi r_0^3\cdot\rho_0}{(M-2\cdot \frac{4}{3}\pi r_0^3\cdot \rho_0)}\right]^{-1/2}.
\ee
Although it is tempting to identify $\frac{4}{3}\pi r_0^3\cdot\rho_0$ as the mass of the vacuum interior to the average radial distance of the elliptical-like orbit, as would be the case in classical physics, one must be careful when calculating volume in the non-Euclidean spacetime of general relativity.

Recall that the functional form of the angular separation between successive like-apsides for a marble orbiting on a warped spandex fabric in the small slope regime is of the form
\be\label{smallsol2}
\Delta{\phi}=220^\circ\left[1+\left(2\alpha(M+\pi r_0^2\cdot\sigma_0)/r_0\right)^{2/3}\right]^{1/2}\left[1+\frac{1}{4}\frac{\pi r_0^2\cdot\sigma_0}{\left(M+\pi r_0^2\cdot\sigma_0\right)}\right]^{-1/2}.
\ee
Comparing the denominators of the last terms in Eqs. (\ref{reldelphi}) and (\ref{smallsol2}), we find that the areal mass density, $\sigma_0$, of the spandex fabric plays the role of a \textit{negative} vacuum energy density, $-\rho_0$, of spacetime.  In each respective theory, the mass of the central object competes for dominance with the mass term associated with the vacuum/spandex fabric.  
As observational evidence implies that we live in a universe with a nonzero (albeit tiny), \textit{positive} vacuum energy density, one can envision a spandex fabric with a negative areal mass density when the conceptual analogy of particle orbits in general relativity and rolling marble orbits on a warped spandex fabric is employed.  This fictional spandex fabric would feel a repulsive force from the uniform gravitational field of the earth and tend to hang \textit{up} towards the ceiling rather than down towards the floor.  
This contribution to the total warping of the spandex surface gives rise to a repulsive force on the orbiting marble, effectively competing with the attractive force of the central mass contribution.  

\vspace{-0.5cm}

\end{appendix}


\begin{thebibliography}{99}
\bibitem{weller}
The theoretical and experimental work presented in this manuscript is based on the year-long senior research project of one of the authors (DW).

\bibitem{embedding}
An embedding diagram is a 2D cylindrically-symmetric surface that has precisely the same spatial curvature as a 2D spatial ``slice" of a respective 4D spacetime.

\bibitem{English}
L. Q. English and A. Mareno, ``Trajectories of rolling marbles on various funnels", Am. J. Phys. {\bf{80}} (11),  996-1000  (2012).

\bibitem{me2}
Chad A. Middleton and Michael Langston, ``Circular orbits on a warped spandex fabric", Am. J. Phys. \textbf{82} (4), 287-294 (2014).

\bibitem{GW}
Gary D. White and Michael Walker, ``The shape of ``the Spandex" and orbits upon its surface", Am. J. Phys. {\bf{70}} (1),  48-52 (2002).

\bibitem{DL}
Don S. Lemons and T.C. Lipscombe, ``Comment on ``The shape of `the Spandex' and orbits upon its surface", by Gary D. White and Michael Walker", Am. J. Phys. {\bf{70}} (10), 1056-1058 (2002).

\bibitem{Nauenberg}
Michael Nauenberg, ``Perturbative approximation for orbits in axially symmetric funnels", Am. J. Phys. \textbf{82} (11), 1047-1052 (2014).

\bibitem{me}
Chad A. Middleton, ``The 2D surfaces that generate Newtonian and general relativistic orbits with small eccentricities", Am. J. Phys. \textbf{83} (7), 608-615 (2015).

\bibitem{apocenter}
The term apsides refers to the maximum and minimum radial distances of an orbiting body.  The apocenter (pericenter) refers to the point on the elliptical orbit that is farthest from (closest to) the central mass.\cite{Marion}

\bibitem{Marion}
Stephen T. Thornton and Jerry B. Marion, \textit{Classical Dynamics of Particles \& Systems} (Harcourt Brace Jovanovich, Inc., 1988).

\bibitem{white}
Gary D. White, ``On trajectories of rolling marbles in cones and other funnels", Am. J. Phys. \textbf{81} (12), 890-898 (2013). 

\bibitem{conic}
The exact solution to Newtonian gravitation is of the form $r(\phi)_{NG}=r_0/(1+\varepsilon\cos\phi)$, where $r_0$ is known historically as half of the latus rectum.  For small eccentricity this expression can be expanded and equates to Eq. (\ref{sol}) to first-order in $\varepsilon$ for $\nu=1$.

\bibitem{Tracker}
For more information about Tracker, or to download, please visit $\langle$http://www.cabrillo.edu/$\sim$dbrown/tracker/$\rangle$

\bibitem{eccentricity}
For an apocenter-pericenter-apocenter orbit, having radii $r_{max,1},\;r_{min},\;$ and $r_{max,2}$, respectively, the average apocenter distance was first calculated through $\overline{r}_{max}=(r_{max,1}+r_{max,2})/2$ and then the eccentricity was calculated through $\varepsilon=(\overline{r}_{max}-r_{min})/(\overline{r}_{max}+r_{min})$.  Likewise, for a pericenter-apocenter-pericenter orbit, having radii $r_{min,1},\;r_{max},\;$ and $r_{min,2}$, respectively, the average pericenter distance was first calculated through $\overline{r}_{min}=(r_{min,1}+r_{min,2})/2$ and then the eccentricity was calculated through $\varepsilon=(r_{max}-\overline{r}_{min})/(r_{max}+\overline{r}_{min})$.

\bibitem{apsides}
In the large slope regime, where an abundance of elliptical-like orbits are generated for a given run, we choose to neglect the angular separation between successive pericenters.  This differs from the small slope regime where we measured the angular separation between successive apocenters and between successive pericenters.

\bibitem{hartle}
James B. Hartle, \textit{Gravity: An Introduction to Einstein's General Relativity} (Addison Wesley, 2003).

\bibitem{carroll}
Sean M. Carrol, \textit{Spacetime and Geometry: An Introduction to General Relativity} (Addison Wesley, 2004).

\bibitem{Stuchlik}
Z. Stuchl\'{i}k and S. Hled\'{i}k, ``Some properties of the Schwarzschild-de Sitter and Schwarzschild-anti-de Sitter spacetimes", Phys. Rev. D {\bf{60}}, 044006 (1999).

\bibitem{Hackmann}
E. Hackmann and C. L\"{a}mmerzahl, ``Geodesic equation in Schwarzschild-(anti-)de Sitter space-times: Analytical solutions and applications", Phys. Rev. D {\bf{78}}, 024035 (2008).

\bibitem{Cruz}
N. Cruz, M. Olivares, and J. Villanueva, ``The geodesic structure of the Schwarzschild Anti-de Sitter black hole", Classical and Quantum Gravity {\bf{22}}, 1167 (2005).


\end{thebibliography}
 \end{document}